# Bibliometrics-based heuristics:

# What is their definition and how can they be studied?

Lutz Bornmann*+ & Sven Hug**

*Administrative Headquarters of the Max Planck Society

Division for Science and Innovation Studies

Hofgartenstr. 8,

80539 Munich, Germany.

Email: bornmann@gv.mpg.de

**Department of Psychology

Binzmühlestrasse 14,

8050 Zürich, Switzerland.

Email: sven.hug@uzh.ch

+ Corresponding author

**Abstract**

When scientists study the phenomena they are interested in, they apply sound methods and base their work on theoretical considerations. In contrast, when the fruits of their research is being evaluated, basic scientific standards do not seem to matter. Instead, simplistic bibliometric indicators (i.e., publications and citation counts) are, paradoxically, both widely used and criticized without any methodological and theoretical framework that would serve to ground both use and critique. Yet, Bornmann and Marewski [1] proposed such a framework recently. They developed bibliometrics-based heuristics (BBHs) based on the *fast-and-frugal heuristics approach* [2] to decision making, in order to conceptually understand and empirically investigate the quantitative evaluation of research as well as to effectively train end-users of bibliometrics (e.g., science managers, scientists). Heuristics are decision strategies that use part of the available information and ignore the rest. By exploiting the statistical structure of task environments, they can aid to make accurate, fast, effortless, and cost-efficient decisions without that trade-offs are incurred. Because of their simplicity, heuristics are easy to understand and communicate, enhancing the transparency of decision processes. In this commentary, we explain several BBHs and discuss how such heuristics can be employed in practice (using the evaluation of applicants for funding programs as one example). Furthermore, we outline why heuristics can perform well, and how they and their fit to task environments can be studied. In pointing to the potential of research on BBHs and to the risks that come with an under-researched, mindless usage of bibliometrics, this commentary contributes to make research evaluation more scientific.

A common practice in research evaluation is to present all available information about an institution, a research group, or a researcher to the decision makers [1]. For example, when the University of Zurich (Switzerland) assessed an academic unit, comprehensive data on organization, resources, research, teaching, and services were compiled and made available to the external peers who evaluated the unit [3]. Such practices are based on the implicit assumption that the more information is used and the more it is processed, the better the assessment and decision will be [4]. However, an emerging body of research questions this assumption. In numerous domains of decision making, ranging from finance and crime to medical diagnosis and sports forecasting, complex, information-greedy procedures can be outperformed by *heuristics* [5]. Heuristics are simple decision strategies that ignore information [5]. In doing so, they can aid to make accurate, fast, effortless, and cost-efficient decisions without that trade-offs are incurred (e.g., effort versus accuracy). Moreover, because they are simple, heuristics allow making transparent decisions that can easily be communicated to and be understood by others.

Bornmann and Marewski [1] transferred the heuristics framework to research evaluation and introduced bibliometrics-based heuristics (BBHs). As other fast-and-frugal heuristics, BBHs are conceived as adaptive judgement strategies that ignore information about some performance aspects (e.g., number of editorial board memberships or scientific prices), thereby allowing quick and robust decisions in research evaluation. In this commentary, we discuss how such heuristics could be defined and empirically studied. We introduce the "science of heuristics" [5] and sketch out what questions could be answered when applying that body of work to evaluative bibliometrics (i.e., research evaluation based on publication and citation counts). We discuss limitations and unsolved problems of the heuristics approach and outline future research on BBHs, pointing to other heuristics and methods to investigate them.

# The science of fast-and-frugal heuristics and bibliometrics

The role of bibliometrics is particularly interesting in evaluation procedures in which peers are informed by indicators. How do peers use indicators in their judgments and how heavily do they rely on them? Could indicators partly or completely replace the judgment of peers? For example, the *EMBO Young Investigator Programme* – aiming at supporting the best young researchers in the life sciences – requires that "applicants must have published at least one last author research paper in an international peer reviewed journal from independent work carried out in their own laboratory" [6]. Do the reviewers count the number of these papers for assessing the applicants and do they take journal metrics into account? Do they look at the *h* index of the applicants (i.e. the number of papers with at least *h* citations)? Should they? Is the predictive accuracy of bibliometric indicators so strong that peer review offers only marginal additional insights?

These and similar questions can be answered through the lens of the *fast-and-frugal heuristics research framework*, originally introduced by Gigerenzer et al. [2] to the cognitive and decision sciences, and extended to multiple other areas since then [e.g., 7,8,9]. Heuristics can produce accurate, fast, effortless or otherwise smart decisions based on little information (hence fast-and-frugal). This has been shown in numerous computer simulations, mathematical analysis, and experiments [10,11]. The assumption is *not* that a given heuristic will yield accurate decisions in all situations. Rather, the art of smart decision-making consists of selecting the heuristic that fits a given decision environment. Indeed, a large body of empirical evidence suggests that people seem to draw from a repertoire (an *adaptive toolbox*) of simplifying, fast-and-frugal heuristics in many domains of decision making, ranging from business and medicine to sports and crime [5]. Bornmann and Marewski [1] discuss in detail that this may also hold true for research evaluation. In particular, the authors introduce the notion of BBHs to explore the potential of the use of heuristics by scientists and evaluators in research evaluation as well as the performance of such heuristics.

BBHs can exploit bibliometric statistics, such as citation or publication counts. In doing so, they may reduce the efforts and time for decision making in research evaluation. Moreover, since BBHs are based on simple computations (e.g., the number of papers which belong to the $x$% most frequently cited papers in the corresponding field), they are easy to understand and apply. The choice of the bibliometric indicators for a particular evaluation should follow guidelines published by experts from the bibliometric community [12,13].

Research on BBHs does not ask whether bibliometric instruments are *generally* or *always* accurate and valid – as has been done in various critical statements on bibliometrics [e.g. 14]. Rather, analogous to the fast-and-frugal heuristics program, research on BBHs takes an ecological perspective and asks in which task environment bibliometric instruments lead to satisfying judgements – and in which not. For example, when it comes to evaluating the research output of different countries it might be possible to rely on metrics only (i.e., only using BBHs); when it comes to assessing single researchers, in turn, BBHs might be better replaced or complemented by an informed peer review process [15].

Key to BBHs and other fast-and-frugal heuristics is the notion of *uncertainty*. Uncertainty is what characterizes decision making in the real world [16]. By definition, under uncertainty, not everything is known and surprises can occur. For example, not all possible courses of action or options, their outcomes, or probabilities of occurrence can be specified [see 17,18] [see 19,20, for a discussion]. Fast-and-frugal heuristics are models for understanding and managing uncertainty. They can perform well even when time, information-processing capacities, and knowledge are limited. Fast-and-frugal heuristics work, because they exploit the statistical structure of information in decision makers' environments and nestle into the workings of basic components of cognition, including the ways in which the perceptual and memory system encode, store, retrieve, and process information [21].

The fast-and-frugal heuristics research program stands in the tradition of Herbert Simon's [22,23] work. Simon stressed that cognition is adapted to the structure of environments. Moreover, he put forward the notion of *bounded rationality* which holds that human information processing capacities are limited. In contrast, the economic maximization models do not assume such bounds, but conceive rational decision making in terms of an exhaustive and complex process, characterized, for instance, by collecting and evaluating all information, weighting each piece of information according to specific criteria, and mathematically integrating that information to come to an 'optimal' solution. Despite Simon's [22,23] critique, the unbounded rationality view still can be found in science and society today and let people think that seemingly rational, exhaustive, and information-greedy procedures always outsmart simple heuristics.

Such maximization thinking might fuel the view that complex research evaluation methods are generally to be preferred over simpler ones (e.g., simple bibliometric indicators). The consequence may be a missed opportunity: like other "mortals" [see 24], research evaluators do not have unlimited information-processing capacity, time, and knowledge. The science of fast-and-frugal heuristics applied to research evaluation might uncover ways of how to deal with such constraints. In the following, we will present two BBHs that could be employed for selecting candidates for a fellowship program (such as the EMBO Young Investigator Programme, see above). Both BBHs are meant to be examples for illustrating the basic principles of heuristics and their potential use in research evaluation. As these two BBHs are candidate models, they need to be subjected to systematic and empirical performance evaluations in given research evaluation environments – a point we will elaborate on further below.

# Models of bibliometrics-based heuristics: Selection of applicants to fellowship programs

The *one-cue BBH* and the *one-reason BBH* could be applied in the pre-selection and final selection of applicants to a fellowship program, such as the EMBO Young Investigator Programme. The two BBHs focus only on two dimensions: *publications* and *citations*. These BBHs implement lexicographic, non-compensatory decision processes, which are defined as follows [25]: "A decision is made lexically when a subject chooses A over B because it is judged to be better along a single, most crucial dimension, without regard to the other 'compensating' virtues that B might have relative to A. Thus, for instance, one would have chosen some restaurant A over B in a lexical fashion if one chose it because it were cheaper and did not care about other trait, like its quality, proximity, level of service, and so on" (p. 8). Lexical decisions are particularly suitable in uncertain situations, such as research evaluation, which is characterized by the lack of knowledge about all consequences of decisions and probabilities of 'correct' decisions [see 26].

Experts, who know which dimensions of a task are most important, seem to make decisions in a non-compensatory way. Betting on just the most important dimension is not a bad thing to do. The non-compensatory *take-the-best heuristic* [27], for instance, has been found to make more accurate predictions than complex information-integration models (e.g., multiple regression) across 20 different task environments from psychology, sociology, demography, economics, health, transportation, biology, and environmental science [see 28,10,11]. Moreover, in many environments, the use of such heuristics can be a response of decision makers to situations where complex and frequent decisions need to be made but capacities and resources are limited.

According to Gigerenzer and Gaissmaier [5], heuristics consist of three building blocks: (1) *search rules* specify where to search for information, (2) *stopping rules* specify the end point of a search, and (3) *decision rules* specify how the final decision should be made. These rules can be flexibly used to shape an assessment process.

**One-cue bibliometrics-based heuristic**

Akin to other heuristics of consideration set generation [e.g., in consumer choice or probabilistic inference; see 29], the search, stopping, and decision rules of the one-cue BBH select, in a first step, the most promising candidates from a pool of applicants. In a second step, a peer review panel assesses the most promising candidates in detail. In pruning down candidates to a smaller *consideration set* [30] of promising ones, this heuristic might reduce the complexity, and, thereby, also the monetary costs of the evaluation procedure.

This elimination process is based on a single indicator which targets an important goal of the funder: research excellence. The three building blocks for the one-cue BBH read as follows: (1) Search all papers published by an applicant in a database that has a good coverage of the applicant's papers and that provides citation data. Find all publications in the database and send the publication list to the applicant for validation. (2) If questionable publications can be either included in or excluded from the applicant's publication list, the list is validated and finalized. Then determine which publications from the list belong to the 10% most frequently cited papers in the corresponding subject category and publication year in the database (i.e., are highly cited papers). (3) Sort the applicants by the number of highly cited papers in descending order and select the top *x*% of the applicants (i.e., those who need to be discussed by the peer review panel). This BBH takes into account the number of publications and their quality (measured in terms of citation impact).

The one-cue BBH can be formalized as follows [the formalizations are based on 31]: Compare the performances of several applicants, $P(1), P(2), P(3), \ldots, P(m)$ using indicator *i*, with $i(1), i(2), i(3), \ldots, i(m)$. Suppose that *x* applicants can be selected. The applicants are sorted by *i* in descending order. If the rank position of *P*(*m*) <= *x*, *P*(*m*) is selected for panel discussion; if *P*(*m*) > *x*, the applicant is not selected.

**One-reason bibliometrics-based heuristic**

One-cue BBHs are special cases of one-reason BBHs. Suppose that there are two applicants who performed very similarly in the past and the experts in the peer review panel have difficulties in selecting one of them for funding. In these situations, random selection has been proposed as selection mechanism [32]. Random selection might prevent selection decisions from biases. One issue of random selection are the expectations of the applicants: applicants assume that decisions are based on scientific or meritocratic criteria only. Scientific and meritocratic selection could be ensured if bibliometric indicators are used as reasons in decision making (e.g., number of highly cited papers, number of collaborating authors, number of single-author publications, number of papers in reputable journals). For example, a one-reason BBH could be formulated to avoid random selection: (1) Select a reason (e.g., number of highly cited papers) and look for the corresponding scores of both applicants. (2) Compare the scores of the two applicants. (3) If they differ substantially, stop and choose the applicant with the better score. (4) If the applicants do not differ, return to the starting point of this loop (step 1) and look for another reason (e.g., number of papers in reputable journals).

Similar to the stopping rule implemented in the take-the-best heuristic, the one-reason BBH bases an inference on the first indicator that *discriminates* between the applicants, that is, on the first indicator for which one applicant has a significantly better score than the other [27]. Different search rules can be implemented. A simple option is to select the indicators from a bibliometric report about applicants randomly one by one [4]. This search rule corresponds to the one implemented in the *minimalist heuristic* [33]. A more elaborated search rule akin to that of the take-the-best heuristic could be used, in which the indicators are ordered by their importance for the specific objectives of the funding agency. If the goals are research excellence (first goal), high degree of collaboration (second goal), and interdisciplinary research (third goal), the selection starts with the number of highly cited papers, followed by the number of collaborating authors, and ends with the number of subject categories to which publications have been assigned.

The decision process of such a one-reason BBH can be formalized as follows [the formalizations are based on 31]. We compare the performances of two applicants: $P(A)$ and $P(B)$. The bibliometric report about $P(A, B)$ contains indicators $i_1, i_2, i_3, \ldots, i_m$. The scores of the indicators on applicant $A$ are symbolized by $i_1(A), i_2(A), i_3(A), \ldots, i_m(A)$ and the scores for applicant $B$ by $i_1(B), i_2(B), i_3(B), \ldots, i_m(B)$. Then, the decision rule of one-reason BBH is as follows:

Infer $P(A) > P(B)$ if and only if $i_m(A) > i_m(B)$,

where $i_n(A) = i_n(B)$ for all $n < m$.

Thus, indicators are inspected one at a time until an indicator is found that has different scores for the two applicants. The applicant with the higher score on this indicator is inferred to have the higher performance.

# Bibliometrics-based heuristics: Beyond the bias view on research evaluation and human decision making

Simon's [22,23] notion of bounded rationality has made it into textbooks. Yet, the ecological stance of his work is often overlooked and, what is worse, limited information processing has become associated with 'bad performance'. Since the 1980s, much empirical work on human decision making and rationality has stressed how mental shortcuts can lead to biases, errors, and fallacious judgments. This influential line of research, developed by Tversky, Kahneman and colleagues [e.g., 34] and known as 'heuristics and biases program', has shaped popular beliefs on heuristics and limited information processing. *Adjustment* and *anchoring*, *availability*, or *representativeness* heuristics largely stand for biases – deviations from classic norms for rationality, defined by Bayes' theorem, the rules of logic, or expected utility maximization. Yet, with very few exceptions, most of these heuristics have never been precisely defined in terms of algorithmic or mathematical models. Instead, vague verbal notions have largely been invoked to explain empirical findings. Much the same holds true for offspring of the heuristics and biases program, notably dual process theories of human reasoning [e.g., 35] that distinguish between a fast but error-prone intuitive (heuristic) system of reasoning and a slow, effortful but more accurate analytic system.

When the fast-and-frugal heuristics program was developed in the 1990s, it has replaced the vague verbal notions of the heuristics and biases approach with algorithmic and mathematical models that can be subjected to both mathematical analysis and computer simulations. It has also replaced the content-blind norms of the heuristics and biases approach with ecological ones: the performance of heuristics must be evaluated with respect to the task environment and the goals of the actor at hand, and not in terms of the fit between heuristic decisions and content-blind norms, such as the rules of logic, Bayes theorem, or expected utility maximization. Hence, performance evaluations of fast-and-frugal-heuristics examine, through mathematical analysis and computer simulations, in which environments heuristics work and in which they fail (i.e., in which environments decision makers would be better off relying on other decision strategies).

As BBHs are fast-and-frugal heuristics, future research on BBHs has (i) to formulate precise algorithmic or mathematical models of BBHs and (ii) to establish ecological benchmarks to evaluate them (i.e., the performance of BBHs with respect to the evaluation task at hand). In this way, BBHs differ from the heuristics and biases approach, which is characterised by vague verbal notions of heuristics and content-blind norms for good decision making.

The following example illustrates the ecological stance of BBHs. In some research fields, the number of papers in 'top-tier' journals is what counts to get a position or to obtain funding. Such a practice contradicts the general, content-blind recommendations of professional bibliometricians [13,12], who would consider this practice as biased, insufficient, or incorrect. In contrast, the BBH approach would ask what (statistical) elements of evaluation environments using such a simple counting decision strategy will lead to desirable and undesirable results and why.

## How the study of BBHs can aid informed judgment

Above, we explained that rather than considering bibliometric indicators based on content-blind norms as 'biased', the study of BBHs would ask what (statistical) elements of evaluation environments using bibliometrics will lead to accurate inferences of future performance. In addition to theoretically grounding the critique of and debate on the usage of bibliometrics in research evaluation, the program proposed by Bornmann and Marewski [1] provides a methodological basis for a debate about values. In particular, BBHs can be designed to predict and understand the performance of a scholar, a department or a university on given criteria, such as the number of future publications or the number of citations. By making explicit how well different indicators and statistics predict such criteria and in which task domains that is the case and in which not, BBHs can shed light on the question whether we, as scientific community or society, think that it is a good idea to base high-stakes decisions (e.g., about professorships, funding of scientific projects) on such performance assessments.

For example, in many countries nowadays political discussions about science funding increasingly revolve around notions such as 'return on investment' or 'accountability for tax payers' money'. For instance, the mission statement of the REF reads: "The assessment provides accountability for public investment in research and produces evidence of the benefits of this investment" [36]. Since BBHs can be formulated as precise algorithmic models, computer simulations can be used to precisely model the results that given certain policies based on bibliometrics might produce. This can aid to examine, a priori, whether the usage of given bibliometric indicators will lead to, potentially, undesirable effects on the aggregate. A typical undesirable effect in science is the Mathew effect named by the Gospel of St Matthew: 'for to all those who have, more will be given' (Matthew 25:29). Merton [37] introduced the term for explaining differences in attention (impact) which eminent and unknown researchers received for similar work. The more frequent attention which eminent researchers receive might lead to a concentration of papers, funding etc.

In a recent study, Bornmann et al. [38] developed a Stata command (h_index) and R package (hindex) to simulate $h$ index and $h_\alpha$ index applications in research evaluation against the backdrop of the BBH framework. Users can apply the command and package to investigate under which conditions the indexes reinforce the Matthew effect. Hirsch [39] proposed the $h_\alpha$ index as a new $h$ index variant: "we define the $h_\alpha$ index of a scientist as the number of papers in the $h$-core of the scientist (i.e. the set of papers that contribute to the $h$ index of the scientist) where this scientist is the $\alpha$-author" (p. 673). The $\alpha$-author is defined "as the author of the paper with the highest $h$-index among all the coauthors" (p. 673). The results of simulations by Bornmann et al. [38] reveal that the $h_\alpha$ index reinforces the Matthew effect.

Future simulation studies might consider funding and hiring decisions. One could set up simulation environments in which different 'agents' (e.g., scientists, units) produce scientific 'output' at a certain rate, with that rate being set to what is observed (e.g., in the discipline, country) empirically. One could then simulate (e.g., funding, hiring) decisions made based on BBHs and examine how, in the long run, the ways in which these decisions are made could lead (or not) to undesired effects on an aggregate level. For instance, in this way it might be possible to precisely predict when using certain indicators will lead to a concentration of funding on only a small number of scientists. The study of BBHs might be helpful in detecting such effects, and, if they are judged to be undesirable (which is a question of values) help to prevent them.

To offer a final example, agent-based simulations with BBHs might also offer grounds for reflecting about other problems the increasing usage of bibliometrics causes in the cosmos of science. One such issue is the gaming of indicators, for instance, through strategic citation or publication efforts undertaking by job applicants (e.g., salami slicing: divide a paper into as many publishable units as possible and then submit them as separate articles [40]; cite your peers' work, expecting that they will cite you in return). With agent-based simulations of BBHs it might be possible to predict and measure in different fields and populations what kind of distributional citation and publication patterns one would expect in a fully 'output-oriented' scientific community.

# Limitations and unsolved problems concerning the science of heuristics and research evaluation

This is not to say that (i) the study of BBHs will be able to solve all problems, or that (ii) the underlying fast-and-frugal heuristics framework is fully free of theoretical and methodological problems itself. Let us elaborate on this point.

In our view any indicator – be it bibliometric or not – risks to represent just a surrogate for the 'real thing', which might not be quantifiable. When asking how important or relevant a scientific finding is for a theory, field, or society, by their very nature, bibliometrics transforms answers to such questions into something that is measurable. However, not all aspects of scientific quality are measurable and quantifiable, and in some cases the relevance of research becomes apparent only with a lag in time (e.g., the importance of a scientific discovery might only be recognized by the community years after publication) [41]. BBHs will likely not aid to stop surrogate science. We join Gigerenzer and Marewski [42] in warning that, just as determining statistical significance and Bayes factors has become a surrogate for good research in certain fields, BBHs that merely capture quantity (e.g., numbers of papers published) might similarly establish quantity as surrogate for quality. For example, the empirical results of Bornmann et al. [43] show that the *h* index mainly focus on the number of papers in a publication set although the *h* index formula also considers the citation impact of the papers.

To counteract such negative effects, a solution might be to design evaluation procedures that include both quantitative and qualitative heuristics. For instance, the one-reason BBHs described in this paper could be complemented by other heuristics that sequentially consider qualitative assessments of peers, rather than bibliometric indicators, as reasons for decision making. These assessments could refer to the following questions: Is the paper the candidate considers to be his/her best piece, also considered to be 'outstanding' by an external reviewer? Is there any evidence that the candidate has built an independent research program? Is the researcher well inter-connected in the field-specific community? Has the researcher published ground-breaking papers? Like BBHs, such review-based heuristics (RBHs) could be studied through the lens of the fast-and-frugal approach as the theoretical and methodological tenets of this approach are agnostic to the type of predictor variable a decision algorithm uses.

It is obvious that no procedure of decision making – including peer-review and fast-frugal-heuristics – is free of error. What matters is to understand when and under which conditions errors occur. If the occurrence of errors is unpredictable, then one has to create environmental conditions that allow to mitigate the error's impact (e.g., safety nets in tenure decisions, such as offering candidates that did not make tenure a follow-up postdoctoral contract). Another challenge faced by the fast-and-frugal heuristic framework and shared by the study of BBHs is that of strategy selection. Approaches to decision making that prescribe (and/or descriptively assume) individuals to choose among a repertoire of decision making procedures as a function of the task environment at hand also need to specify normative (and/or descriptive) models of the choice process. The fast-and-frugal heuristics program has developed some models for strategy selection [44,31]. Corresponding models need to be developed for BBHs and should ideally be included in evaluation guidelines.

## Conclusion

When scientists study the phenomena they are interested in, they apply sound methods and base their research on theoretical considerations. In contrast, when the fruits of their research is being evaluated, basic scientific standards do not seem to matter. Yet, when evaluating the fruits of scientific research, a split-brain mentality seems to reign: simplistic bibliometric indicators (i.e., publication and citation counts) are widely used and, at the same time, widely criticized, but methodological and theoretical frameworks to ground both the widespread use and critique are scarce. Pointing towards this gap, Bornmann and Marewski [1] recently proposed a framework to conceptually understand, empirically study, and effectively train the use of bibliometric indicators. With this commentary, we mean to both introduce this framework to a broader audience and to call for corresponding empirical work to be carried out. If we know that researchers and science managers employ simple bibliometric shortcuts (i.e., BBHs) to reach decisions in research evaluation, one has to investigate how such shortcuts are being used and how successful they are in different evaluative settings.

To conclude, with this commentary, we neither advocate nor condemn bibliometrics for research evaluation. However, as bibliometrics is often criticized for being reductionist compared to peer review approaches in the context of research evaluation, we think that it is time to take the bull by the horns and to ground the debate on the use of bibliometric indicators in a conceptual framework.